\documentclass[fleqn,10pt]{wlscirep}

\usepackage{setspace}
\title{Uncovering large inconsistencies between machine learning derived gridded settlement datasets}

\author[1,*]{Vedran Sekara}
\author[2]{Andrea Martini}
\author[3]{Manuel Garcia-Herranz}
\author[3]{Do-Hyung Kim}

\affil[1]{Department of Computer Science, IT University of Copenhagen, Copenhagen, Denmark}
\affil[2]{United Nations Children's Fund, Latin America and Caribbean Regional Office, Panama City, Panama}
\affil[3]{United Nations Children’s Fund, New York, NY, USA}

\affil[*]{Corresponding author, vsek@itu.dk}

\begin{abstract}
High-resolution human settlement maps provide detailed delineations of where people live and are vital for scientific and practical purposes, such as rapid disaster response, allocation of humanitarian resources, and international development.
The increased availability of high-resolution satellite imagery, combined with powerful techniques from machine learning and artificial intelligence, has spurred the creation of a wealth of settlement datasets.
However, the precise agreement and alignment between these datasets is not known. 
Here we quantify the overlap of high-resolution settlement map for 42 African countries developed by Google (Open Buildings), Meta (High Resolution Population Maps) and GRID3 (Geo-Referenced Infrastructure and Demographic Data for Development).
Across all studied countries we find large disagreement between datasets on how much area is considered settled.
We demonstrate that there are considerable geographic and socio-economic factors at play and build a machine learning model to predict for which areas datasets disagree. 
It it vital to understand the shortcomings of AI derived high-resolution settlement layers as international organizations, governments, and NGOs are already experimenting with incorporating these into programmatic work.
As such, we anticipate our work to be a starting point for more critical and detailed analyses of AI derived datasets for humanitarian, planning, policy, and scientific purposes.

\end{abstract}
\begin{document}

\flushbottom
\maketitle

\thispagestyle{empty}

% ------------------- INTRODUCTION -------------------

\section*{Introduction}

A large range of applications from humanitarian response~\cite{lu2012predictability,acosta2020quantifying}, urban and infrastructure planning~\cite{ratti2006mobile}, epidemic preparedness~\cite{linard2012large} to environmental science~\cite{o2010global}, require detailed maps of where people live.
However, it can be hard to obtain accurate and timely data~\cite{deville2014dynamic}.
Traditionally, accurate maps of human settlements have been derived from census data or representative household surveys.
Although both sources of information provide rich and irreplaceable data, they can be expensive and time consuming.
For instance, the US 2020 national census is estimated to have cost up to \$14.2 billion~\cite{gaogov}.
Further, certain areas might not even be surveyable due to e.g. conflicts or natural disasters.
Because of these limitations, surveys are only performed once every 5 to 10 or 15 years depending on the country, but rapidly changing contexts, e.g. migration, rapid urbanization, humanitarian crises, and pandemics, can quickly make data outdated.

In response to these limitations developmental agencies, non-governmental organizations (NGOs), academics, and governments have explored new technologies and complementary approaches to estimate population densities~\cite{dabalen2016mobile,ureport,lloyd2017high}.
New opportunities have been opened up by increased availability, and better global coverage, of high-resolution satellite imagery, with resolutions down to centimeter level~\cite{wardrop2018spatially,maxar}.
Combined with advances in computation and AI powered image recognition techniques, high-resolution population datasets derived from satellite images have in the recent years become readily available~\cite{tiecke2017mapping,sirko2021continental}.
These gridded population estimates provide detailed maps of where people live~\cite{dobson2000landscan,doxsey2015taking,stevens2015disaggregating,ghs-pop-2015,fries2021measuring,leasure2020national}.
Additionally, some of these datasets also provide insights on the distribution of specific populations, including the number of children under five, the number of women, as well as elderly populations~\cite{tatem2013millennium,alegana2015fine,pezzulo2017sub}.
As such, these population datasets are frequently being used for policy work~\cite{bongaarts2011population} and to monitor the progress of reaching the sustainable development goals~\cite{tuholske2021implications}.

There are two fundamentally different approaches to building high-resolution population estimates~\cite{wardrop2018spatially}, called the top-down and bottom-up methods.
A comprehensive review of the strengths and weaknesses of these methods can be found in Leyk et al. 2019~\cite{leyk2019spatial}. 
The top-down approach uses a weighted collection of geospatial covariates, such as nighttime light, land cover, etc., and a dasymetric mapping to spatially distribute census level data and create gridded population estimates.
The bottom-up approach combines micro-census data with geospatial covariates to build a statistical model which estimates the population for unsurveyed areas.
When it comes to high-resolution population datasets, both approaches have been used.
Today there exists many high-resolution population datasoures which researchers and practitioners can freely choose from~\cite{fries2021measuring}.
For instance, Meta's (previously Facebook) \textit{High Resolution Population Density Maps}~\cite{FB-high-res-pop} make use of a top-down approach to distribute population numbers using a high-resolution settlement layer as covariate, while \textit{WorldPop's} high-resolution population estimate for Guinea uses a bottom-up approach~\cite{leasure-2021}.

\begin{figure}[!htp]
\centering
\includegraphics[width=0.6\linewidth]{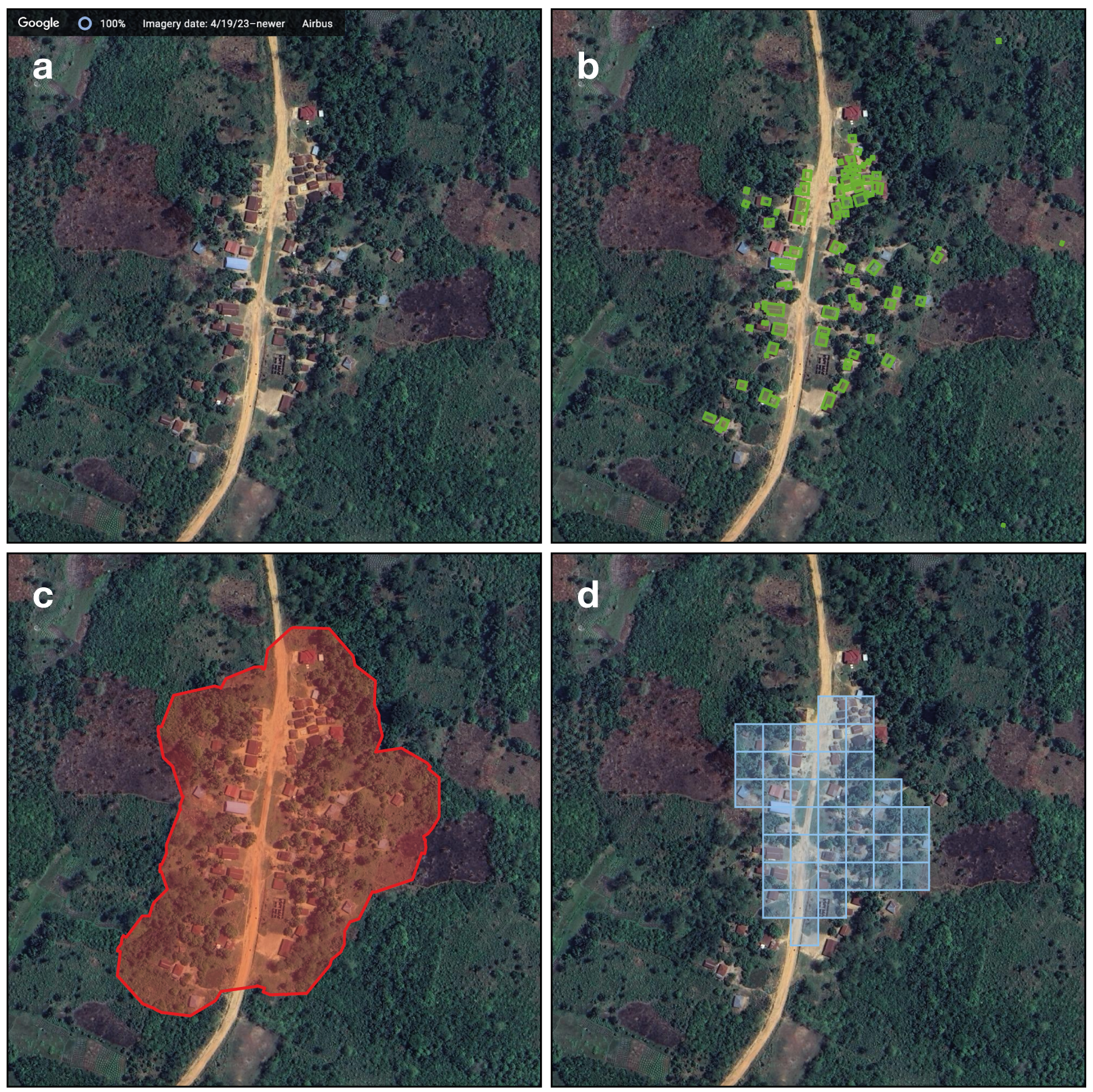}
\caption{\textbf{Illustration of the different settlement datasets}. \textbf{a}, Satellite imagery of houses in the settlement \textit{Pindegumahun} located on the northern outskirts of the city \textit{Bo} in Sierra Leone. \textbf{b}, The Open Buildings (OB) dataset, developed by Google~\cite{sirko2021continental}. Settlement data is provided in csv files containing outlines of buildings (green outline), which are derived from high-resolution satellite imagery. The figure shows the actual inferred OB settlements. There are buildings which are not detected, and there are slight offsets between settlements and the satellite image; this is due to images stemming from two different periods (see Methods for more details). \textbf{c}, The Geo-Referenced Infrastructure and Demographic Data for Development (GRID3) dataset provides settlement data as vector files (i.e. polygons)~\cite{GRID3}. Polygons represent the extent of settled areas (red shared area). \textbf{d}, The High Resolution Settlement Layer (HRSL) developed by Meta~\cite{tiecke2017mapping} is a raster dataset that provides settlement at a resolution of 1 arc-second, approximately 30$\times$30 m at equator (see Methods), here illustrated by blue cells.}
\label{fig:fig1}
\end{figure}

Independent of the type of approach, whether it is top-down or bottom-up, a key geospatial covariate is the human settlement layer.
The settlement layer informs statistical models of where there are human settlements (i.e. regions settled by humans), to either distribute population into (top-down), or from where to start (bottom-up).
In this sense, high-resolution population estimates will only be as good as their underlying settlement layer.
In fact, a recent study concluded that population estimates which are based remote sensing are more informative because they are better able to capture uninhabited areas~\cite{fries2021measuring}. 
While there have been efforts in understanding how different population estimates compare~\cite{leyk2019spatial,archila2020pixel,fries2021measuring,tuholske2021implications}, there has been little focus on quantifying the overlap and mismatch of the various high-resolution settlement datasets (which population estimates are based on).
As such, our study looks into how well existing settlement datasets align, whether they are consistent across countries, and whether regional characteristics such as poverty and development play a role in their level of overlap.
As we expect settlement layers to play an increasingly important role in policy making and international development, we focus on quantifying the inconsistencies across three of the most popular population datasets and their settlement layers. 
This includes: 1) the \textit{Geo-Referenced Infrastructure and Demographic Data for Development} (GRID3)~\cite{GRID3}, 2) \textit{High Resolution Settlement Layers} (HRSL) developed by Meta~\cite{tiecke2017mapping} and the Center for International Earth Science Information Network at Columbia University, and 3) the recently released \textit{Open Buildings} (OB) dataset by Google~\cite{sirko2021continental} (see Fig.~\ref{fig:fig1}, and Methods and Materials for a detailed description of the datasets).
Here we investigate how well these three settlement datasets agree across multiple scales, from country level, down to 100$\times$100 m cells.
A recent study compared building footprints across Africa using similar spatial resolutions~\cite{chamberlain2024building}.
We quantify the overlap, compare it to the socio-economic status of areas, and build a predictive model to understand the causes of if.
While we are interested in quantifying the global agreement between settlement datasets, our study is limited to the African continent as these datasets predominantly exists for African nations (see SI Fig S1 for an overview).

\section*{Comparing settlement datasets}

As datasets contain settlement information in different formats, spatial extents, and at varying geospatial resolutions, we standardize them by performing two steps.
First, if a datasets comes with population information we transform it into a binary raster by truncating values, such that zeroes indicate not-settled areas and ones denote settled areas.
We do this because we are not interested in population densities, rather we are interested in knowing where sophisticated machine learning and artificial intelligence models have identified human settlement.
Secondly, we correct for the different geospatial resolutions and alignments of the datasets by upscaling them to a single raster with 3 arc seconds resolution, which roughly approximately 100 $\times$ 100 m cell-sizes at equator. 
See SI Sec. S1 for more details on the alignment procedure.

\begin{figure}[!htbp]
\centering
\includegraphics[width=\linewidth]{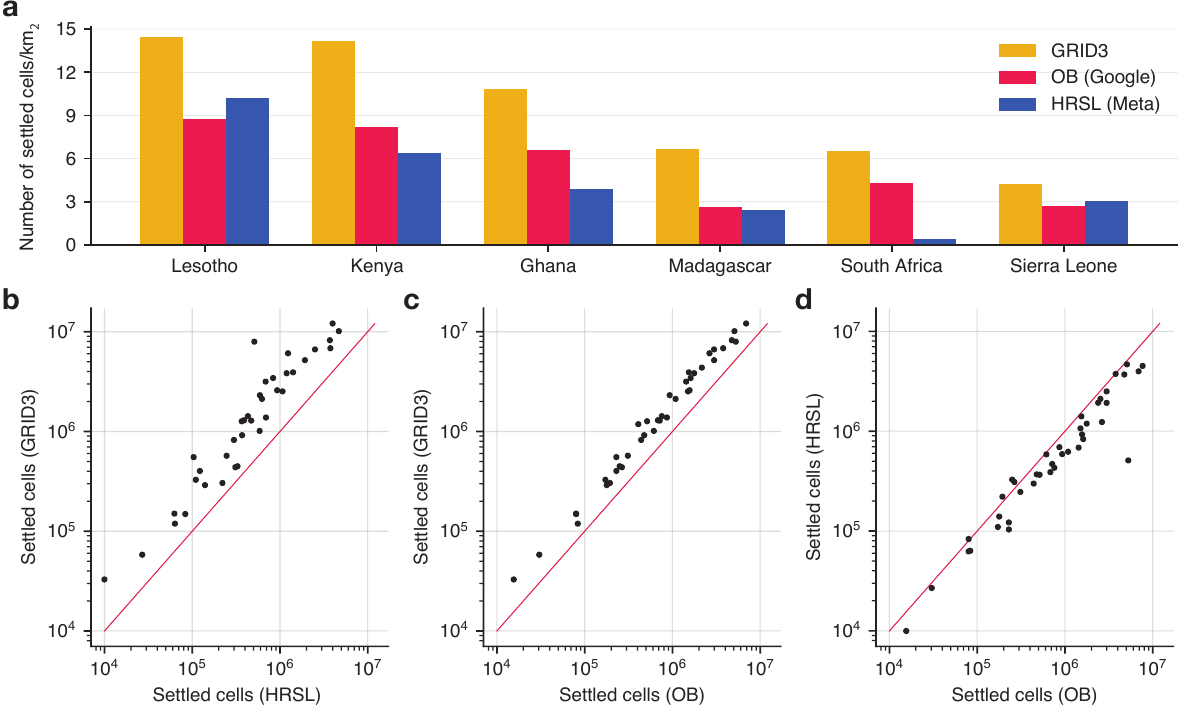}
\caption{\textbf{Comparison of the number of settled cells across datasources}. \textbf{a}, There are large differences in how many settled cells datasets contain per country, shown here for a small sample of countries. Due to differences in country areas we normalized the number of settled cells by the surface area of a country. Countries are sorted according to cells/km$^2$ for GRID3. \textbf{b}, Comparison between the raw number of cells in GRID3 and HRSL shows that GRID3 consistently contains more settled cells (correlation $R=0.90; p\ll10^{-6}$). Each black dot signifies a country. \textbf{c}, GRID3 also consistently contains more settled cells than OB (correlation $R=0.99; p\ll10^{-6}$). \textbf{d}, The relationship between HRSL and OB is less clear ((correlation $R=0.89; p\ll10^{-6}$). For a large majority of countries OB contains more cells, but there are also countries for which HRSL has identified more settled cells.
}
\label{fig:fig2}
\end{figure}

As a first indicator of how well settlement dataset agree we measure the number of settled cells per square kilometer.
Fig.~\ref{fig:fig2} shows data for seven countries, for which all 3 datasets are available (see SI Fig. S1 for information on data coverage), revealing there can be large differences in the number of settled cells. 
Overall, GRID3 datasets always contain the highest number of settled cells per km$^2$. 
This is, in part, due to its choice of settlement definition (i.e. total settlement extents) which covers more of the surrounding areas compared the other datasets (Fig.~\ref{fig:fig1}b-d).
However, when we look at the gap between GRID3 and OB and HRSL, and the gap between OB and HRSL we already find large disagreements.
In some cases the gap between GRID3 and the other datasts can be relatively modest.
For example, for Sierra Leone, GRID3 has $4.2$ settled cells/km$^2$, while HRSL and OB respectively have $3.1$ and $2.7$.
This corresponds to roughly $26\%$ difference for HRSL and $36\%$ for OB compared to GRID3.
For other countries the gap can be more pronounced.
For instance, for Madagascar HRSL and OB contain between 2.4-2.6 cells/km$^2$, while GRID3 has 6.7 cells/km$^2$, i.e. more than double. 
For South Africa the situation is different, here HRSL contains 0.4 cells/km$^2$, while OB has 4.3, and GRID3 6.5 cells/km$^2$.
To quantify the overall differences we look at the correlation between the dataset in terms of settled cells. 
GRID3 consistently contains more cells than either HRSL and OB (Fig.~\ref{fig:fig2}b,c), while the relationship between HRSL and OB is less consistent.
For a majority of countries OB contains more settled cells, but for some countries HRSL does dataset contains more cells.
Although Fig.~\ref{fig:fig2}b-d suggests there are consistent scaling factors between the datasets, the large differences in number of settled cells will undoubtedly have considerable impacts on gridded population numbers, as population estimates will have to be distributed over strongly varying number of cells.

\begin{figure}[!htbp]
\centering
\includegraphics[width=\linewidth]{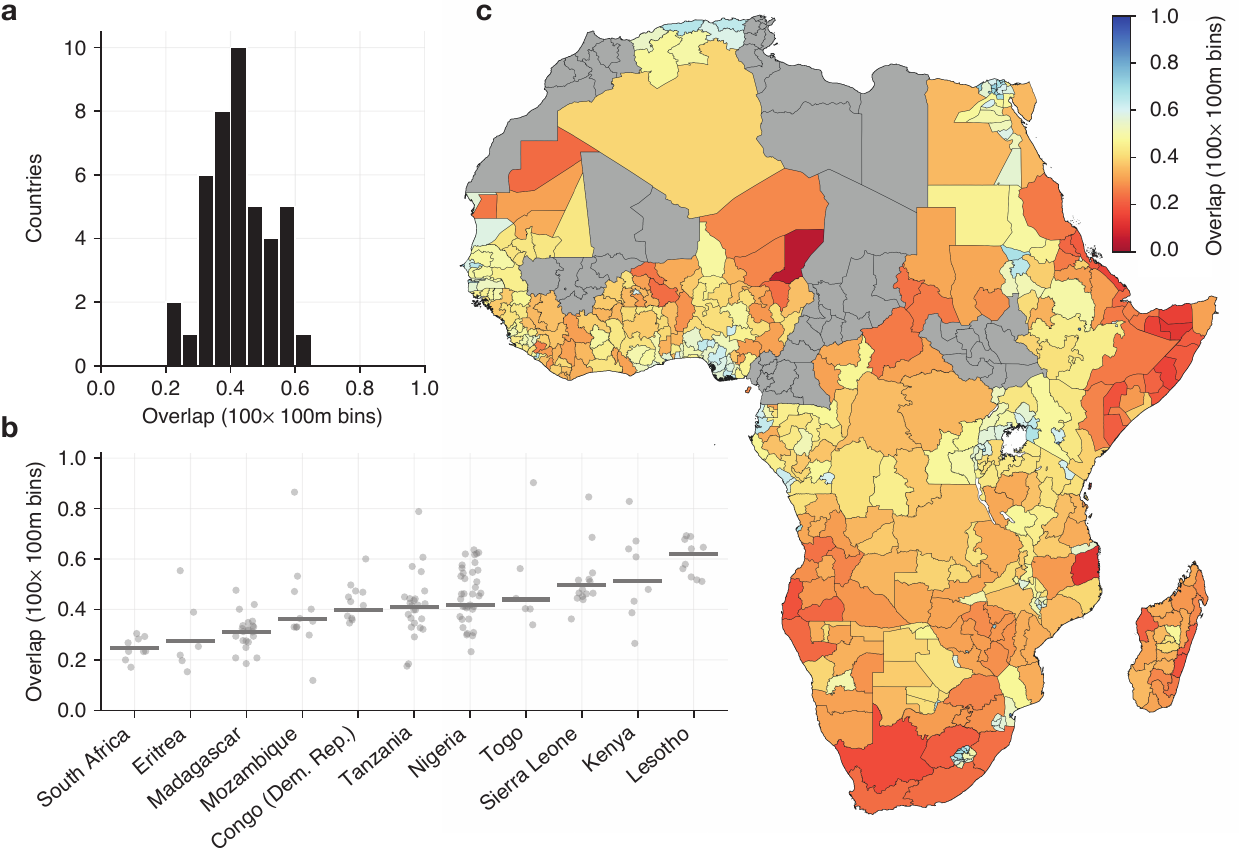}
\caption{\textbf{Large variations in overlap on sub-national level between settlement datasets}. \textbf{a}, Average overlap between datasets for 42 countries. High-resolution settlement dataset agree, on average, on only 42.6\% of all cells. \textbf{b}, National level statistics can hide a lot of nuance. Full lines denote overlap on a country level, while circles denote overlap on a regional level (admin 1). \textbf{c}, Spatial distribution of agreement  between settlement datasets for administrative regions in 42 countries. Overlap is calculated based on $100\times100$ m settlement rasters, and summarized for each region.} 
\label{fig:fig3}
\end{figure}

Fig.~\ref{fig:fig2} only compares the number of settled cells between the datasets, not how well the different datasets actually align.
To quantify agreement between datasets we calculate the overlap as $\theta = J(X,Y)$ (see Methods), where $J(X,Y)$ is the Jaccard index between two rasters (upscaled to the same spatial resolution).
We use the Jaccard index as it only calculates overlap between settled cells (where people live); it does not include unsettled areas (where people do not live) into the calculation.
If two datasets completely overlap $\theta = 1$, and if they completely disagree, i.e. no overlap, then $\theta = 0$.
As we (for most countries) have three datasets we need a method to compare the total mismatch across multiple datasets.
We do this by calculating the average overlap between datasets as $\theta = 1 / N\sum_i^N \theta_i$. 
For instance, for a country where three settlement rasters are available ($X$, $Y$, and $Z$) we calculate overlap as $\theta = 1/3 (\theta_{XY} + \theta_{XZ} + \theta_{YZ})$.
Fig.~\ref{fig:fig3}a shows the observed distribution of overlap for 42 countries in Africa where at least two datasets are available.
The average overlap across all countries is $0.43$, but there is a relatively high variability between countries (standard deviation of $0.096$). 
The lowest overlap (lowest agreement) is for Somalia ($0.23$), while the highest agreement is for Lesotho ($0.62$).
Overall, this reveals that settlement overlap is surprisingly low.
However, the disagreements in numbers of settled between datasets (Fig.~\ref{fig:fig2}) can play a role in lowering the observed overlap.
For instance, if one dataset has 100 settled cells, while another dataset contains 1000 cells, the highest achievable theoretical overlap is $\theta_{\text{upper limit}} = 0.1$.
To quantify the size of this effect we calculate the upper limit of overlap for each country ($\theta_{\text{upper limit}}$, see Methods), which corresponds to a situation where datasets perfectly align, i.e. where the smaller raster is entirely contained in the larger one. 
We do this to distinguish how much of the mismatch originates from differences in dataset sizes versus how much arises from datasets not agreeing which areas are settled.
For instance, for Rwanda $\theta_{\text{observed}} = 0.57$ while the maximally possible overlap is $\theta_{\text{upper limit}} = 0.64$.
This means the observed overlap is around $89\%$ of the maximally possible overlap, so a large part of the mismatch for Rwanda arises from settlement datasets identifying different numbers of settled cells.
For other countries the upper limit is considerably higher than the observed overlap (see SI Sec. S2).
For example, for Madagascar the observed overlap is around half of the maximally possible overlap ($\theta_{\text{observed}} = 0.31$, $\theta_{\text{upper limit}} = 0.56$), suggesting that mismatch is not only caused by differing numbers of settled cells, but also by AI generated settlement datasets fundamentally disagreeing on which areas are settled.
In general, it is unclear whether the mismatch between settlement rasters stems from data providers using different definitions of what a settlement is, differences in AI methodologies, satellite imagery from different periods, or whether it is caused by them using different confidence values for settlements (the HRSL dataset does not contain any associated confidence values for settled areas, while GRID3 and OB do).
Nonetheless, the fact remains, there are considerable disagreements, even when the size of datasets are accounted for. 

Disagreements between rasters might also be caused by other effects such as noise, or offset issues between datasets.
To test this, we downsample settlement resolutions from $\sim 100\times100$m cells to larger sizes, up to $\sim 3\times3$km cells (see SI section S3) and calculate the overlap for all countries as a function of cell size (see SI Fig. S5).
When downsampling, we say a larger cell is settled if any of the smaller $100\times100$m cells within it are settled.
For two of the most densely populated countries in Africa, Burundi and Rwanda, we find that doubling the cell size by a factor of two, to $\sim200$m resolution, drastically increases overlap by up to 19 percent point (from 0.45 to 0.64 for Burundi and 0.57 to 0.75 for Rwanda).
This indicates there are local disagreement in settlement datasets which are 'smoothed over' by larger cell dimensions.
In urban areas these disagreement are often caused due to settlements having many contiguous buildings without clear delineations between them~\cite{sirko2021continental}.
Unfortunately this relationship does not hold for all countries, for instance, for Djibuti and Gabon (SI Fig. S5) large disagreements persist even for large cell sizes ($\sim3\times3$km, with $\theta_{100\textrm{m}} = 0.37$ and $\theta_{3\textrm{km}} = 0.39$).
The heterogeneity across countries demonstrates that we need to dig deeper into the factors behind disagreements.

So far we have only focused on county level statistics (Fig.~\ref{fig:fig3}a), but to better understand what drives disagreement we look at disagreement on sub-national level.
Using Global Data Lab~\cite{GDLv4} shapefiles we split settlement datasets into administrative regions (admin 1 level) and calculate overlap for individual regions.
Fig.~\ref{fig:fig3}b shows the overlap for 10 countries, with lines denoting country-level overlap while grey circles show overlap for individual regions. 
As seen in the figure, there can be large variations in agreement, even within countries.
For instance, in Mozambique the capital region, Maputo Cidade, has an overlap value of $\theta = 0.87$, while for Cabo Delgado, the poorest administrative region in the country, the agreement between datasets is only $\theta = 0.12$.
Mozambique is not an outlier, similar patterns can be found in other countries.
Fig.~\ref{fig:fig3}c shows the regional disagreement for all admin 1 regions of the 42 countries where at least 2 datasets are available.
It illustrates that inconsistencies between high-resolution settlement datasets are not limited to one particular geographical location, rather, high levels of inconsistency appear to be present across the African continent.

\section*{Understanding which factors contribute to mismatch}
To uncover the factors behind dataset mismatch, we start by comparing dataset overlap to the Human Development Index~\cite{undp-hdi} (HDI).
The HDI which is a composite index developed by the the United Nations to estimate the development of a region, not only in terms of economic growth alone, but across three key dimensions of human development: life expectancy, schooling, and standard of living.
HDI is available on administrative region level (admin 1) and is bounded between zero and one, where zero indicates a minimal development.
Comparing, across all countries, HDI to dataset overlap (on admin region level 1) reveals a positive association (Pearson $r = 0.41$, $p\ll  10^{-6}$, see SI Sec S4), meaning settlement datasets tend to have higher overlap for regions with higher HDI (see SI Fig. S7).
Unfortunately, HDI is not a high-resolution measure; we can only use it to uncover relationships for large regions.

To get a more accurate understanding of which factors play a role in mismatch, we study at the problem at a finer granular level (cells of $100$m resolution) and build a machine learning model to predict overlap. 
As features for the model we use high-resolution data from three sources. 1) the Relative Wealth Index (RWI) and its associated error values~\cite{chi2022microestimates} which contains micro-estimates of wealth for all low- and middle-income countries. 2) Nightlight intensity captured globally by satellites and harmonized by Li et al.~\cite{li2020harmonized}. And 3) the Global Human Settlement Layer~\cite{florczyk2019ghsl} which contain information about settlement types (e.g. low density rural, rural, suburban, urban, etc.) and allows us to measure how often disagreements occur depending on the settlement type of the region.

The prediction problem can be approached from multiple angles.
One can merge data across countries and build one global model, or build individual models for each country.
We have chosen to do both, but in the main manuscript we only report results from the global model (see SI Sec S6 for results on country models).
We start by first creating a joint dataset by concatenating data from all countries for which all three datasets exist (37 countries in total, see SI Sec. S5 for the full list).
The joint dataset contains around 108 million datapoints, with each datapoint representing a geographical cell of $100\times100$m. 
Only cells which, at least one, of the datasets considers settled are included in the data.
Cells which all datasets consider unsettled are not included in the analysis.
(We have chosen to do this as we are interested in modeling overlap, not whether cells are settled or not.)
We treat it as a binary classification problem where the negative class (label 0) denotes situations where one or more of the datasets disagree on whether a cell is settled, while the positive class (label 1) represents cases where all three datasets agree.
Overall, $24.2\%$ (26M) of datapoints are labeled as class 1, while the remainder are class 0. 
(For class distributions for individual countries see SI Fig S9.)
As model we use an L2 regularized logistic regression to classify whether datasets agree or disagree. 
A logistic regression is a relatively simple model, its strengths lie in it being transparent and easy to interpret.
For example, model weights give us the odds ratio, determining how a change in one variable (keeping all other variables constant) affects the odds of datasets agreeing.
We use the F1-score as model scoring function and evaluation metric as it is robust to class imbalances, however, we have chosen to also report balanced accuracy, as it is easier to understand.

\begin{figure}[!htbp]
\centering
\includegraphics[width=\linewidth]{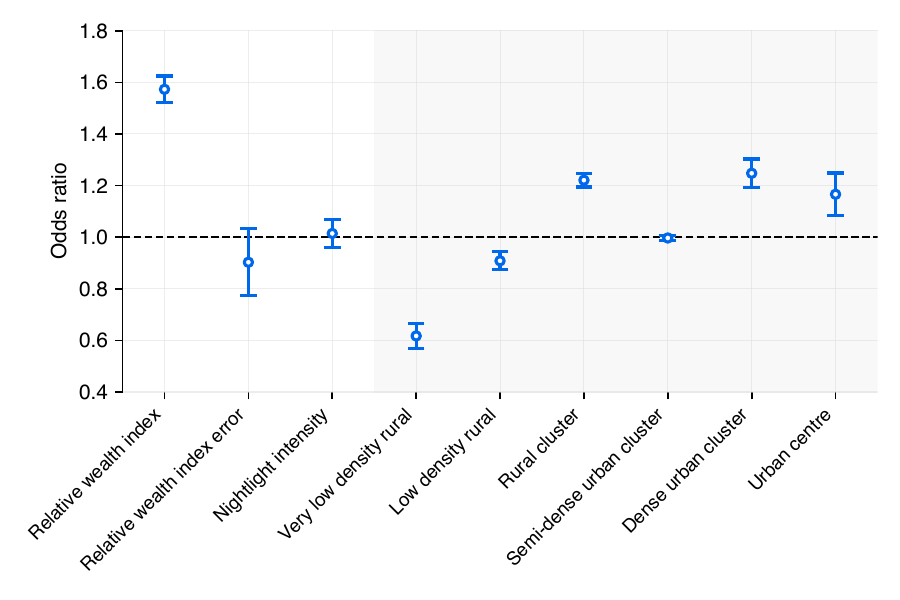}
\caption{\textbf{Determining which factors influence overlap between high-resolution settlement datasets.} Coefficient estimates of the model, excluding intercept, including 95\% confidence intervals estimated from 100 bootstrap samples. Confidence intervals are relatively narrow compared to the overall coefficient weights. (Fig SI S12 shows more clearly the distribution of variance for coefficients.) The grey shaded area indicates variables related to the settlement type.}
\label{fig:fig4}
\end{figure}

Our model training uses a nested cross validation setup to ensure we avoid bias when evaluating performance~\cite{cawley2010over} (i.e. we avoid using the same cross-validation splits to both tune and select a models, as this often leads to optimistically biased evaluations).
Both inner and outer loops use 5-fold cross validation, where the inner loop selects hyperparameters (in our case, the regularization), and the outer loop evaluates the generalizability and performance of models.
The inner loop uses a stratified 5-fold scheme to preserve the percentage of samples for each class, while the outer loop used a group 5-fold setup to avoid information leakage by ensuring datapoints from a country do not appear in both test and train data.
For each outer loop we save the best model.
Overall, the average F1 score is 0.423 [0.276; 0.49] and balanced accuracy is 0.615 [0.563; 0.654] (numbers in brackets show the total range of results, from the lowest to the highest performing models).
Our goal is not to get a perfect prediction, rather we want to explain which variables are indicative of settlement datasets agreeing or disagreeing.
Fig.~\ref{fig:fig4} shows the average weights of the models, plotted as odds ratios.
As the Global Human Settlement Layer provides categorical feature, we transform the seven categories (see methods) into six features using one-hot encoding with the \textit{Suburban or per-urban grid cells} as reference point.
This means that odd ratios for other settlement variables are relative to suburban cells. 
RWI, RWI error and nightlight intensity are all numerical features that are harmonized across countries, we do not pre-process them in any way (other than standardizing them during the cross validation loop, see Fig S10 for data distributions). 
From Fig.~\ref{fig:fig4} we observe that the variable with the highest odds ratio is RWI, where an increase in relative wealth changes the odds of datasets agreeing by a factor of 1.57 [1.49; 1.62].
The estimated error in RWI and the nightlight intensity do not have any noteworthy impact (their odds ratios are within to 1).
When it comes to settlement features, the odds for settlement datasets agreeing in very low density rural areas are lower by a factor of 0.62 [0.58; 0.69] compared to suburban or pre-urban areas.
Similarly, low density rural areas have odds 0.91 [0.86; 0.95] for dataset agreement compared to suburban areas.
Rural clusters, dense urban clusters, and urban centers have all odds higher than one; respectively 1.22 [1.18; 1.25], 1.25 [1.16; 1.29], and 1.17 [1.07; 1.24] relative to suburban areas.
Semi-dense urban clusters have identical odds to suburban areas.
Initially it might seem counter-intuitive that rural clusters have odds above one, but given that this variable represents densely populated rural areas, with at least 300 inhabitants per square km\cite{florczyk2019ghsl}, it is not surprising that settlement datasets are likely to overlap in such regions.
In summary, our analysis shows that settlement dataset are more likely to agree in higher developed, wealthier, and more densely populated areas. 

To better understand the role population density plays in the overlap of settlement datasets we divide cells into two categories (low- and high-
density areas), where cells that are categorized as rural centers or above are considered high population density areas (the rest are considered low density). 
Looking at overlap rates using this binary categorization we find that gridded settlement datasets disagree significantly more in low-density areas (see SI Table S3).
For some countries, such as Rwanda, we find a modest ratio of disagreement, where we are 1.22 times more likely, in high population density areas, to observe an overlap between settlement datasets, compared to low density regions.
However, for a majority of countries this gap is larger.
For example, for Angola and C\^{o}te d’Ivoire gridded settlement datasets are respectively 3 and 2.5 times more likely to agree in high population density areas (see SI Table S2).
Put differently, error rates for rural low population density areas are 3 to 2.5 times higher than for more urbanized places.

\section*{Discussion}
Satellite earth observations provide affordable, fast, and timely information on human settlements and population estimates.
As a result, a lot of work has focused on building complex machine learning models to estimate various population characteristics~\cite{sirko2021continental,GRID3,tiecke2017mapping}.
However, despite drastic improvements these models, and the datasets they produce, still have many limitations.
To externally evaluate model performance, their ability to generalize, and audit for bias one requires access to the machine learning models and the underlying satellite imagery. 
Unfortunately, the models and training data are rarely open source.
As such, to evaluate the limitation of machine learning models used for population estimation and settlement identification we focus on their outputs.
Here, we investigate to which degree AI produced high-resolution human settlement layers agree for countries in Africa. 
Our choice of evaluating the overlap between settlement layers, rather than gridded population estimates, is based on two factors.
First, populations datasets are built on top of settlement data.
In fact, case-studies have shown that population estimates can vary widely depending on the population dataset of choice~\cite{archila2020pixel,tuholske2021implications}, however, there is a gap in the literature when it comes to comparing settlement layers.
Second, the lack of ground truth (e.g. micro census estimates) on a global scale makes it infeasible to perform large-scale comparisons of population data.

Our analysis presents three key findings: first, there are large inconsistencies between settlement datasets from HRSL, GRID3 and OB.
For some countries the difference in number of cells identified as being settled can be immense, with some dataset having ten times more settled cells than others (see Fig.~\ref{fig:fig2}).
This mismatch will undoubtedly impact any population mapping methodology which uses these datsets as base layer.
Put differently, imperfections in the underlying settlement data will propagate to population estimates.

Second, even when correcting for the size of settlement dataset (i.e. controlling for the number of cells) we still observe large disagreements.
For some countries disagreements disappear when cell resolution is reduced, e.g. to $200\times200$m cells, highlighting that mismatch mainly stems from minor inconsistencies.
However, for other countries, this is not the case, here disagreement persist even for large cell-sizes (up to $3\times3$km cells), indicating that machine learning models fundamentally disagree on which cells contain human settlements.
Unfortunately countries do not only fall into one of these two categories, it is a board spectrum.
If high-resolution settlement and population datasets are to be used robustly to monitor the progress towards the United Nations sustainable development goals~\cite{SDG}, policy making, and for responding to sudden onset emergencies, these issues need to be understood in more detail, and addressed.
Future work could look into uncovering the optimal, or 'natural', resolutions at which settlement datasets can be aggregated to maximize overlap.

Third, we built a model to estimate where mismatch between gridded settlement datasets occurs, finding that poorer, less developed, and less densely populated places have the highest rates of mismatch.
This does not come as a surprise, as population density and urban/rural factors have previously been determined to affect gridded population estimates~\cite{archila2020pixel}, and impact the mismatch in building footprints~\cite{chamberlain2024building}.
We trained one model by merging data from multiple countries, however, it is also possible to build individual models for each country (see SI Sec. S6).
Although similar conclusions can be drawn from the country-models there are some variation across which variables are informative for each country. 
Future work could focus on understanding these differences in more detail, and on exploring, and building separate models for various geo-spatial contexts (e.g. urban/rural).
Some of the inconsistencies in dataset overlap can stem from studies using satellite imagery from different periods.
Although we tried to minimize the issue by focusing on recently released dataset which use satellite imagery from comparable years, this can still be a potential major source of disagreement.
Unfortunately, it has been impossible to determine the exact temporal information for satellite imagery used to derive settlement layers.
Temporal information, such as the date of satellite imagery used to estimate settlement information, should be added to these datasets.

It is vital to understand the shortcomings of AI derived high-resolution settlement and population layers as international organizations, governments, and NGOs are already experimenting with incorporating these into programmatic work, including resource allocation and policy making\cite{blumenstock2020machine,aiken2022machine,unicef-pop}.
Taking UNICEF's recently developed Children's Climate Risk Index as an example; it employs gridded population data (and its underlying settlement layer) to gauge the child population impacted by various climate risks~\cite{rees2021climate}.
If using different population datasets can lead to contrasting outcomes in the index, this will significantly influence policy-making processes and priority setting. 
The fundamental disagreements between datasets, which we have demonstrated, raises the question: how should international organization, academics, NGOs and local governments know which map is the best for their purpose?
For populations numbers, it has been argued that a single datasource perspective should be avoided when tracking developmental progress~\cite{tuholske2021implications}, as depending on which dataset is used the population estimates can be up to $\pm$ 1 million people.
A number which can have a large effect on a humanitarian responses efforts, e.g. it can result in under or over-allocation of resources.
Similar issues have been raised for land cover maps~\cite{kerner2023accurate}, where no single map can be considered 'best' when evaluated across multiple metrics and countries. 
Based on our results it is impossible for us to say which settlement dataset is globally optimal to use. 
One approach that can be beneficial, and help reduce potential errors, is to create an ensemble by combining multiple settlement layers. 
This has previously been proposed for population estimates, in order to capture uncertainty and provide intervals for population numbers~\cite{fries2021measuring,tuholske2021implications}.
Similar solutions for settlement layers should be explored.
However, before creating ensembles of settlement datasets it is vital to first understand individual model accuracy across a range of context (urban/rural, country, biome, etc.).
For instance, if the machine learning models, which produced the settlement maps, consistently have low accuracy and confidence intervals for rural and less developed areas, merging their predictions will not necessary improve settlement maps for these places; it might just introduce more noise.
Future work should focus on standardizing reporting of performance metrics across multiple geo-spatial contexts~\cite{sekara2023machine}, and on boosting model performance for the lowest-performing contexts (e.g. less developed regions, urban/rural, specific countries).
Ensemble learning is further hindered by gridded settlement datasets using different confidence scores (or likelihood scores) for 'defining' settlements.
For example, the OB dataset recommends setting a threshold and using only shapes with confidence scores of at least $0.7$, GRID3 recommends a threshold of $0.5$, while HRSL data does not contain any confidence scores for the underlying settlement data.
In our study we have used the recommended confidence scores from each dataset. 
For practitioners knowing which combination of datasets to use and how to set the proper thresholds is difficult.
Standardizing definitions on what a `human settled' cell entails and developing standards for settlement layers will alleviate many of these inconsistencies and help build more robust settlement maps.

% ------------------- METHODS -------------------

\section*{Methods and Materials}

\noindent \textbf{The Open Buildings (OB) dataset}~\cite{sirko2021continental} (v3) released by Google is available for 43 countries in mainland Africa (see SI Fig. S1c). Data is released in csv format and contains outlines of identified buildings along with a confidence score, denoting how confident the AI model is in each individual prediction (i.e. the models' confidence that an identified shape in fact is a building). To remove locations of low confidence, we adopt the definition of high-confidence scores from the original paper~\cite{sirko2021continental}, and only include shapes that have a confidence score of at least $0.7$. Model inference for OB was carried out during May 2023, however, it is unclear which period satellite images are from. While the Google team tried to include the most recent images, in some cases, the most recent image for a location was several years old, or not available.

\noindent \textbf{The Geo-Referenced Infrastructure and Demographic Data for Development (GRID3)}~\cite{GRID3} settlement extent dataset is given in the form of a geodatabase of polygons per country. Each polygon denotes the extent of a settlement, or where the presence of buildings detected from satellite imagery suggest that there is likely a human settlement. We use the most up-to-date version of GRID3 (v2). Settlements are inferred from building footprints derived by the Ecopia AI using a feature extraction algorithm on Maxar satellite imagery captured between 2009 and 2021. It is intended to be used for government and humanitarian planning with data being available for 43 countries, see SI Fig. S1a. In addition, each polygon is associated with a probability of it being a false positive. The probability distributions can vary considerably across countries (see SI Fig. S3). We focus only on polygons which have a false positive probability lower than $0.4$ (i.e. above 60\% probability of being correctly labeled).

\noindent \textbf{High Resolution Settlement Layer.} Developed by Meta and the Center for International Earth Science Information Network at Columbia University~\cite{tiecke2017mapping}, the data can be downloaded from the Humanitarian Data Exchange~\cite{hdx-fb-link}. Data for each country is given in a raster format with resolution of 1 arc-seconds (approximately 30x30 m) and exists for 40 countries (see SI SI Fig. S1b). The data originally contains population estimates, where population numbers for each administrative region are uniformly distributed across all identified settled grids. To extract the settlement layer we binarize each raster, see SI Sec 1 for information on threshold values. We use version 2 of the data, which contains the most up-to-date settlement information. However, in contrast to version 1 of the dataset version 2 rasters are unmasked, meaning they contain population information for points outside the country. As such, we mask rasters using shapefiles from the Global Data Lab~\cite{GDLv4}.

\noindent \textbf{The Global Human Settlement Layer~\cite{florczyk2019ghsl}}, produced by the European Commission, provides information on the type of settlement (e.g. low density rural, rural, suburban. urban, dense urban, urban center) on a $1\times1$ km geospatial resolution. We use this data to understand where the GRID3, Facebook, and Google high-resolution settlement layers disagree, and quantify whether it is more likely to happen in rural or urban areas. This dataset is given in Mollweide projection and has been reprojected to align with the coordinate system of the other rasters (WGS84). The raster divides human settlements into 8 types. This includes: urban centre, dense urban cluster, semi-dense urban cluster, suburban or per-urban grid cells, rural cluster, low density rural, very low density rural, and water cells. A cell is denoted as water cell if it has 0.5 share covered by permanent surface water and is not populated nor built. However, in a few cases the settlement datasets do indicate that some of these cells are settled. As such, we have chose to include water cells in the study, but have chosen to merge them with the class 'very low density rural'. As such, we keep the information from water cells, but denote it as very low density rural.

\noindent \textbf{The Relative Wealth Index~\cite{chi2022microestimates},} provides information on relative standards of living within countries. It is developed in collaboration with Meta by training machine learning algorithms on multiple and heterogeneous datasets including satellite imagery, mobile phone data, connectivity data, and topographic maps. It is provided in raster format with 2.4km resolution. 

\noindent \textbf{The Human Development Index~\cite{undp-hdi}} is developed by the United Nations Development Programme. It provided a composite index developed of human development measured across multiple dimensions, including: life expectancy, schooling, and standard of living. We use the 2021 version (which is also the most recently available) of the subnational Human Development Database developed by Smits and Permanyer~\cite{smits2019subnational}.

\noindent \textbf{Nightlight intensity,} data is from Li et al.~\cite{li2020harmonized}. The data is harmonized and converted time serie nightlight intensity data captured by VIIRS (Visible Infrared Imaging Radiometer Suite) for 2014-2021. We use the 2021 version (v7) of the dataset. The spatial resolution is 30 arc-seconds, which approximately corresponds to 1km resolution.

\noindent \textbf{Estimating overlap.} 
Settlement data is stored in so-called raster files which can be thought of as geo-referenced matrices.
When two rasters are given on the same spatial resolution is is possible to estimate mismatch as the Jaccard overlap between two matrices, $X$ and $Y$.
For boolean matrices (with 0 or 1 values) this is given as $J(X,Y) = |X\cap Y| / |X\cup Y|$.

\noindent \textbf{Estimating the upper limits on overlap.}
For scenarios where two rasters align perfectly, i.e. one is fully contained in the other, overlap can be calculated as $\theta_{\text{upper limit}} = \min(|X|,|Y|) / \max(|X|,|Y|)$, where $|\cdot|$ indicates the number of settled cells in a raster. We use this to understand the gap between the overlap we observe versus what is maximally possible.

\noindent \textbf{Predictive model.}
We use a logistic regression to classify individual 100m resolution cells (or pixels), where we denote a cell to have the label 0 if at least two of the datasets agree on whether the cell is settled, and 1 if the datasets agree. We only include cells which at least one of the dataset consider settled; areas which all dataset consider to be unsettled are not included. For model training we use a nested cross validation setup, where both inner and outer loops use 5-fold cross validation. The inner loop uses a stratified 5-fold scheme to preserve the percentage of samples for each class, while the outer loop used a group 5-fold setup, where data from countries is grouped together, to avoid information leakage by ensuring datapoints from a country do not appear in both test and train data. For each outer fold we build one model. Error bars are estimated across the 5 outer folds.

\bibliography{sample}

\end{document}